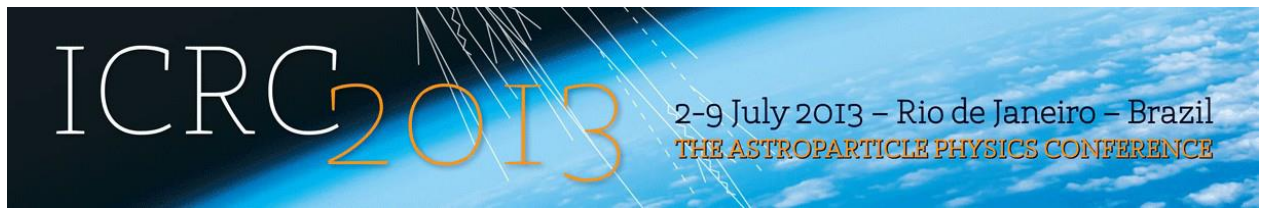

# The Space-Based Gamma-Ray Telescope GAMMA-400 and Its Scientific Goals

A.M. Galper[1,2], O. Adriani[3], R.L. Aptekar[4], I.V. Arkhangelskaja[2], A.I. Arkhangelskiy[2], G.A. Avanesov[5], L. Bergström[6], E.A. Bogomolov[4], M. Boezio[7], V. Bonvicini[7], K.A. Boyarchuk[8], V.A. Dogiel[1], Yu.V. Gusakov[1], M.I. Fradkin[1], Ch. Fuglesang[9], B.I. Hnatyk[10], V.A. Kachanov[11], V.V. Kadilin[2], V.A. Kaplin[2], M.D. Kheymits[2], V. Korepanov[12], J. Larsson[9], A.A. Leonov[2], F. Longo[7], P. Maestro[13], P. Marrocchesi[13], V.V. Mikhailov[2], E. Mocchiutti[7], A.A. Moiseev[14], N. Mori[3], I. Moskalenko[15], P.Yu. Naumov[2], P. Papini[3], M. Pearce[9], P. Picozza[16], M.F. Runtso[2], F. Ryde[9], R. Sparvoli[16], P. Spillantini[3], S.I. Suchkov[1], M. Tavani[17], N.P. Topchiev[1*], A. Vacchi[7], E. Vannuccini[3], G.I. Vasiliev[4], Yu.T. Yurkin[2], N. Zampa[7], V.N. Zarikashvili[18], V.G. Zverev[2]

[1] Lebedev Physical Institute, Russian Academy of Sciences, Moscow, Russia
[2] National Research Nuclear University MEPhI, Moscow, Russia
[3] Istituto Nazionale di Fisica Nucleare, Sezione di Firenze and Physics Department of University of Florence, Firenze, Italy
[4] Ioffe Physical Technical Institute, Russian Academy of Sciences, St. Petersburg, Russia
[5] Space Research Institute, Russian Academy of Sciences, Moscow, Russia
[6] Stockholm University, Department of Physics; and the Oskar Klein Centre, AlbaNova University Center, Stockholm, Sweden
[7] Istituto Nazionale di Fisica Nucleare, Sezione di Trieste, Trieste, Italy
[8] Research Institute for Electromechanics, Istra, Moscow region, Russia
[9] KTH Royal Institute of Technology, Department of Physics; and the Oskar Klein Centre, AlbaNova University Center, Stockholm, Sweden
[10] Taras Shevchenko National University of Kyiv, Ukraine
[11] Institute for High Energy Physics, Protvino, Moscow region, Russia
[12] Lviv Center of Institute of Space Research, Lviv, Ukraine
[13] Istituto Nazionale di Fisica Nucleare, Sezione di Pisa and Physics Department of University os Siena, Siena, Italy
[14] NASA Goddard Space Flight Center and CRESST/University of Maryland, Greenbelt, Maryland, USA
[15] Hansen Experimental Physics Laboratory and Kavli Institute for Particle Astrophysics and Cosmology, Stanford University, Stanford, USA
[16] Istituto Nazionale di Fisica Nucleare, Sezione di Roma 2 and Physics Department of University of Rome "Tor Vergata", Rome, Italy
[17] Istituto Nazionale di Astrofisica – IASF and Physics Department of University of Rome "Tor Vergata", Rome, Italy
[18] Pushkov Institute of Terrestrial Magnetism, Ionosphere, and Radiowave Propagation, Troitsk, Moscow region, Russia

**Abstract:** The design of the new space-based gamma-ray telescope GAMMA-400 is presented. GAMMA-400 is optimized for the energy 100 GeV with the best parameters: the angular resolution ~0.01º, the energy resolution ~1%, and the proton rejection factor ~$10^6$, but is able to measure gamma-ray and electron + positron fluxes in the energy range from 100 MeV to 10 TeV. GAMMA-400 is aimed to a broad range of science topics, such as search for signatures of dark matter, studies of Galactic and extragalactic gamma-ray sources, Galactic and extragalactic diffuse emission, gamma-ray bursts, as well as high-precision measurements of spectra of cosmic-ray electrons + positrons, and nuclei.

**Keywords**: gamma-ray astronomy, gamma-ray telescope, dark matter, electrons, positrons, nuclei

---

[*] Corresponding author, e-mail: tnp51@yandex.ru

## 1. Introduction

The present time of the development of high-energy gamma-ray astronomy outside the Earth's atmosphere is possible to characterize as the Fermi-LAT era [1]. Indeed, the basic information and basic scientific and methodological results are associated with the Fermi-LAT flight. The total number of gamma-ray sources has reached two thousand. Practically, all basic astrophysical objects have appeared as gamma-ray sources. Now, one can classify in detail pulsars, active galactic nuclei, and other objects. A specific feature in the energy spectrum of the high-energy gamma-ray emission from the Galactic center has revealed [2-4]. Obviously, the next important step in the development of astronomy outside the Earth's atmosphere and understanding the nature of the processes occurring in the active variable astrophysical objects, such as the Galactic center, the Cygnus constellation, extended sources, unidentified sources from the Fermi-LAT catalog, the majority of which are located near the Galactic plane, will be obtaining the results by high-energy gamma-ray telescopes with higher angular and energy resolutions than the Fermi-LAT telescope.

Simultaneously with improving the physical characteristics, the signal/background ratio considerably increases that is fundamental to resolve high-energy gamma-ray lines, which can arise when annihilating or decaying the hypothetical dark matter particles.

It is this possibility of studying the dark matter noted at the end of the last century by the Nobel laureate Academician Vitaly Ginzburg, which initiated the GAMMA-400 project in Russia [5, 6]. Within the framework of this project, which has become international, the precision gamma-ray telescope with high physical and technical characteristics is designed [7, 8].

## 2. Scientific goals

The GAMMA-400 main scientific goals are: study of the origin of the dark matter by means of gamma-ray astronomy; precise measurements of Galactic and extragalactic discrete astrophysical sources; research of high-energy gamma-ray bursts; research of high energy electron + positron fluxes; research of high-energy nuclei fluxes.

## 3. GAMMA-400 gamma-ray telescope

GAMMA-400 is optimized for the energy 100 GeV with the best parameters: the angular resolution ~0.01°, the energy resolution ~1%, and the proton rejection factor ~$10^6$, but is able to measure gamma-ray and electron + positron fluxes in the energy range from 100 MeV to 10 TeV. The GAMMA-400 effective area is ~5000 cm$^2$ at $E_\gamma > 1$ GeV, the total mass is 4100 kg, the power consumption is ~2000 W, and a telemetry downlink capability is 100 GB/day. Together with GAMMA-400, the space observatory will include two star sensors for determining the GAMMA-400 axes with accuracy of approximately 5", two magnetometers, and the KONUS-FG gamma-ray burst monitor.

The GAMMA-400 physical scheme is shown in Fig. 1. GAMMA-400 consists of scintillation anticoincidence top and lateral detectors AC, converter-tracker C with 10 layers of double *(x, y)* silicon strip coordinate detectors (pitch 0.1 mm) interleaved with tungsten conversion foils, scintillation detectors S1 and S2 of time-of-flight system TOF, scintillation detectors S3 and S4, lateral detectors LD, neutron detector ND, and calorimeter from two parts CC1 and CC2. The imaging calorimeter CC1 consists of 2 layers of double (x, y) silicon strip coordinate detectors (pitch 0.1 mm) with CsI (Tl) crystals, and the electromagnetic calorimeter CC2 consists of CsI (Tl) cubic crystals. The total converter-tracker thickness is ~1$X_0$ ($X_0$ is the radiation length). The thickness of CC1 and CC2 is 2$X_0$ and 23$X_0$, respectively. The total calorimeter thickness is 25$X_0$ and 1.2$\lambda_0$ ($\lambda_0$ is nuclear interaction length) in the vertical direction, 54$X_0$ and 2.5$\lambda_0$ in the lateral direction.

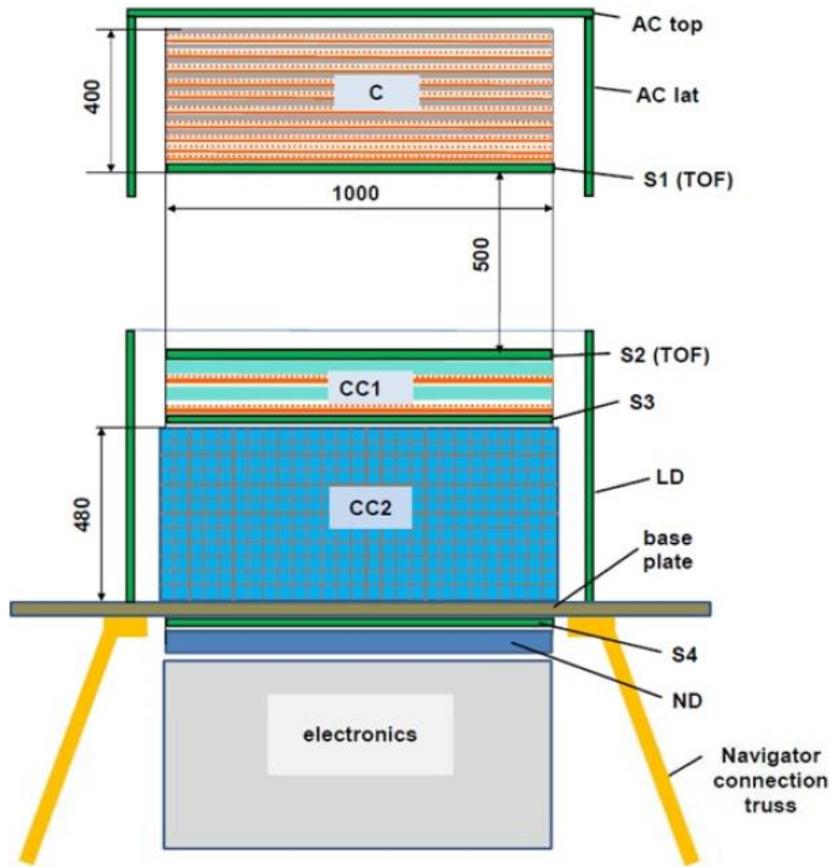

**Figure 1**: GAMMA-400 physical scheme.

Gamma rays are detected by the absence of a signal in AC, while electrons (positrons) are detected by the presence of a signal in AC, when moving downward and from lateral directions. Using the calorimeter with thickness ~$25X_0$ extends the particle measurable energy range up to 10 TeV and increases the gamma-ray telescope energy resolution up to ~1% at energies more than 10 GeV. The energy dependence of the GAMMA-400 energy resolution for incident gamma-rays was simulated using Monte Carlo techniques and is shown in Fig. 2 along with the same dependence for the Fermi-LAT [9] for comparison. It is seen that in the energy range from 10 GeV to ~10 TeV the energy resolution is ~1%, which is extremely important for resolving the gamma-ray lines from the annihilation or decay of the dark matter particles.

High angular resolution is achieved by determining the conversion point in a multilayer converter-tracker and the reconstruction of the shower axis in CC1. This method allows us to obtain the high angular resolution ~0.01° at energies more than 100 GeV (Fig. 3) and enables to locate precisely the source of the gamma-ray lines.

High-energy incident particles create a backsplash (upward moving products of the shower) in the calorimeter. To prevent the detection of the backsplash particles in the AC (thereby creating a self-veto), we use the method of separation of incident and backsplash particles in the AC by the time-of-flight method along with the segmentation method.

The proton rejection factor ~$10^6$ (critical parameter for the background rejection) will be achieved by using the calorimeter and the neutron detector together with other instrument subsystems.

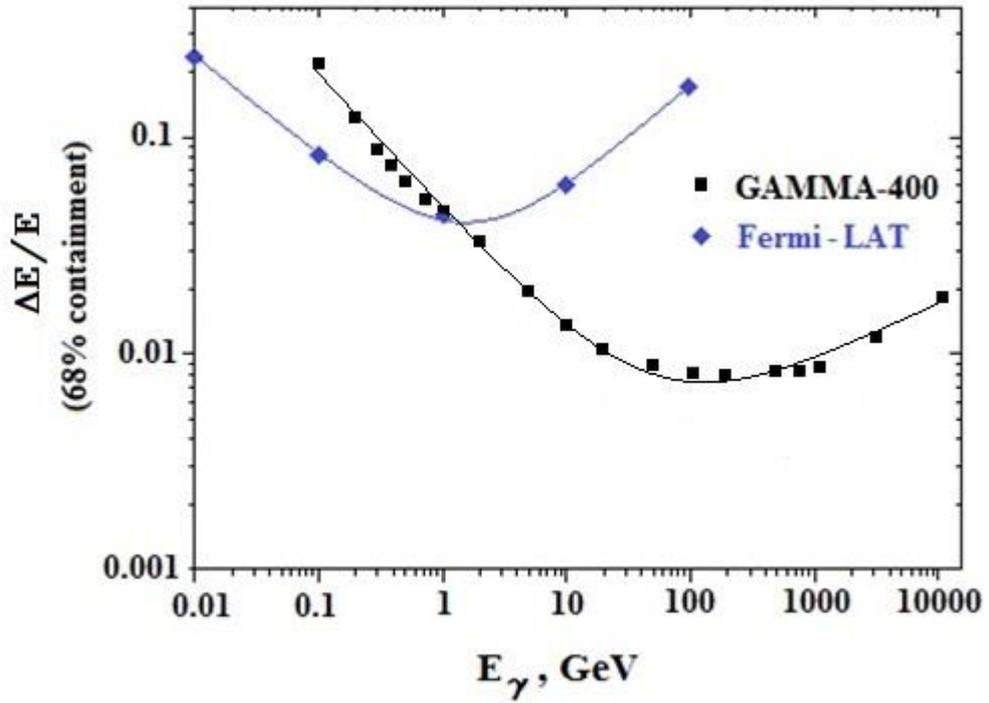

**Figure 2**: Energy resolution vs. energy for normal incidence for Fermi-LAT [9] and GAMMA-400.

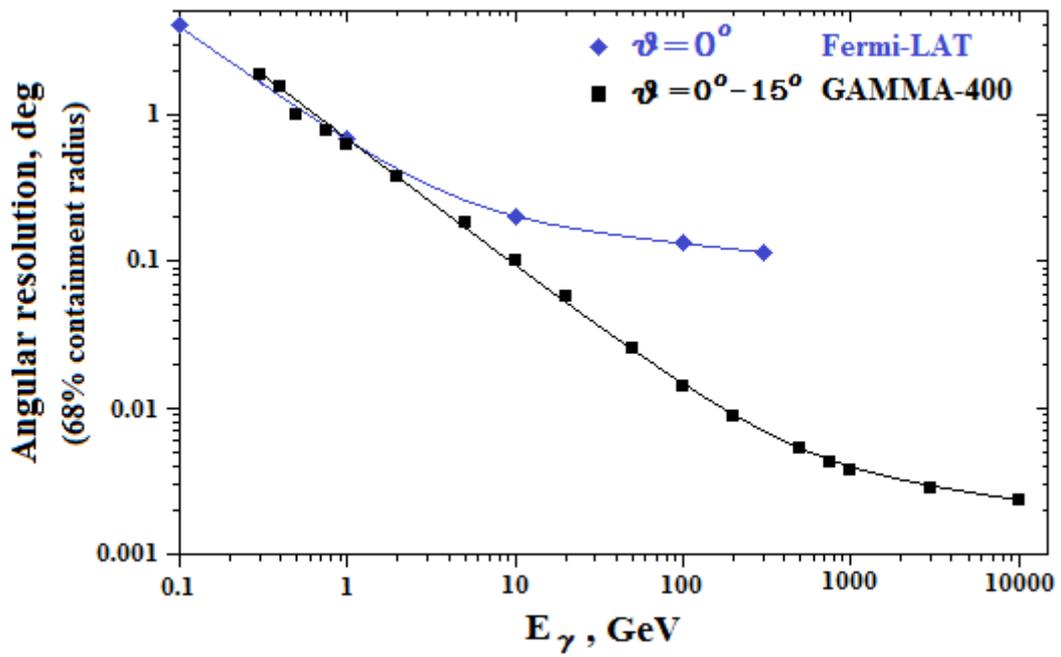

**Figure 3**: Angular resolution vs. energy for Fermi-LAT [9] (for normal incidence) and GAMMA-400 (for θ=0°-15°).

In Table 1, the comparison of the main parameters for GAMMA-400 and Fermi-LAT [1] is shown.

**Table 1**. Comparison of the main parameters for GAMMA-400 and Fermi-LAT [1]

|  | **Fermi-LAT** | **GAMMA-400** |
|---|---|---|
| Orbit | circular, 565 km | high-elliptical, 500-300 000 km |
| Energy range | 20 MeV - 300 GeV | 100 MeV – 10 000 GeV |
| Effective area ($E_\gamma > 1$ GeV) | ~8000 cm$^2$ | ~5000 cm$^2$ |
| Coordinate detectors | Si strips with pitch 0.23 mm | Si strips with pitch 0.1 mm |
| Angular resolution ($E_\gamma > 100$ GeV) | ~0.1° | ~0.01° |
| Calorimeter - thickness | CsI ~8.5$X_0$ | CsI(Tl)+Si strips ~25$X_0$ |
| Energy resolution ($E_\gamma > 100$ GeV) | ~10% | ~1% |
| Proton rejection coefficient | ~$10^4$ | ~$10^6$ |
| Mass, kg | 2800 | 4100 |
| Telemetry downlink capability, GB/day | 15 | 100 |

## 4. Spacecraft

The GAMMA-400 space observatory will be installed on the Navigator space service platform produced by Lavochkin Association and launched by the more powerful Proton-M rocket with the Briz booster. The initial high-elliptical orbital parameters are: an apogee of 300 000 km, a perigee of 500 km, and an inclination of 51.4°. The orbit period will be 7 days. After approximately 230 days GAMMA-400 will leave the Earth's radiation belts and the orbit will change from highly elliptical to approximately circular with median altitude of ~150 000 km.

We are planning to use long-term observations of Galaxy, first of all, Galactic center, Cygnus constellation, Crab, and other extended and variable Galactic and extragalactic sources.

## 5. Conclusion

The GAMMA-400 project is included in the Russian Federal Space Program and funded by the Russian Space Agency. The launch of the GAMMA-400 space observatory is planned in 2018-2019. The expected mission duration is longer than 7 years.

**Acknowledgment**: This work was supported by the Space Council of the Russian Academy of Sciences and the Russian Space Agency.